\PassOptionsToPackage{table}{xcolor}
\documentclass[sigconf]{acmart}
\setcitestyle{square,sort,comma,numbers}

\usepackage[utf8]{inputenc} % allow utf-8 input
\usepackage[T1]{fontenc}    % use 8-bit T1 fonts
\usepackage{hyperref}       % hyperlinks
\usepackage{url}            % simple URL typesetting
\usepackage{booktabs}       % professional-quality tables
\usepackage{amsfonts}       % blackboard math symbols
\usepackage[show]{notes-alt}
\usepackage{nicefrac}       % compact symbols for 1/2, etc.
\usepackage{microtype}      % microtypography
\usepackage{xcolor} % colors
\usepackage[table]{xcolor}

\usepackage{amstext}
\usepackage{amsmath}
\usepackage{enumitem}
\usepackage{algorithm}
\usepackage{algpseudocode}
\usepackage{dblfloatfix}
\usepackage{multirow}
\usepackage{wrapfig}
\usepackage{tikz}
\usepackage{pgfplots}
\usepackage{subcaption}
\usepackage{pgfplotstable}
\usetikzlibrary{patterns}

\newcommand{\idnt}{\phantom{ - }}
\newcommand{\sys}[1]{\textsc{Gar}\def\temp{#1}\ifx\temp\empty{}\else\raisebox{-.4ex}{\scriptsize #1}\fi}
\newcommand{\gar}{\textsc{Gar}}
\newcommand{\sgar}[1]{\textsc{Sgar}\def\temp{#1}\ifx\temp\empty{}\else\raisebox{-.4ex}{\scriptsize#1}\fi}

\newcommand{\sysbm}[1]{\sys{}${}_{BM25}$}
\newcommand{\quamsys}[1]{\textsc{Quam}\def\temp{#1}\ifx\temp\empty{}\else\raisebox{-.4ex}{\scriptsize #1}\fi}

\newcommand{\cerberussys}[1]{\textsc{Ore}\def\temp{#1}\ifx\temp\empty{}\else\raisebox{-.4ex}{\scriptsize #1}\fi}
\newcommand{\cerberuhalfssys}[1]{\textsc{CerberusHalf}\def\temp{#1}\ifx\temp\empty{}\else\raisebox{-.4ex}{\scriptsize #1}\fi}

\newcommand{\quam}{\textsc{Quam}}

\newcommand{\laff}{\textsc{Laff}}

\newcommand{\cerberus}{\textsc{ORE}}

\newcommand{\argmaxm}[1]{%
  \ifthenelse{\isempty{#1}}%
    {\overset{m}{\argmax}}% if #1 is empty
    {\underset{#1}{\overset{m}{\argmax}}\, }% if #1 is not empty
}
\newcommand{\argminm}[1]{%
  \ifthenelse{\isempty{#1}}%
    {\overset{m}{\argmin}}% if #1 is empty
    {\underset{#1}{\overset{m}{\argmin}}\, }% if #1 is not empty
}

\usepackage{pifont}
\newcommand{\cmark}{\ding{51}}%

\author{Mandeep Rathee}
\affiliation{%
  \institution{L3S Research Center}
  \city{Hannover}
  \country{Germany}  
}
\email{rathee@l3s.de}
\author{Venktesh V}
\affiliation{%
    \institution{Delft University of Technology (TU~Delft)}
    \city{Delft}
    \country{The Netherlands}
}
\email{v.viswanathan-1@tudelft.nl}

\author{Sean MacAvaney}
\affiliation{%
  \institution{University of Glasgow}
  \city{Glasgow}
  \country{United Kingdom}
}
\email{sean.macavaney@glasgow.ac.uk}
\author{Avishek Anand}
\affiliation{%
    \institution{Delft University of Technology (TU~Delft)}
    \city{Delft}
    \country{The Netherlands}
}
\email{avishek.anand@tudelft.nl}

\begin{document}
% \title{Online Relevance Estimation for resource-efficient Document Ranking}
% \title{CERBERUS: A Multi Arms Bandit Adaptive Retrieval Approach }

\title{Breaking the Lens of the Telescope: Online Relevance Estimation over Large Retrieval Sets}

\begin{abstract}

\end{abstract}
%    <concept>
%        <concept_id>10002951.10003317</concept_id>
%        <concept_desc>Information systems~Information retrieval</concept_desc>
%        <concept_significance>500</concept_significance>
%        </concept>
%  </ccs2012>

% \ccsdesc[500]{Information systems~Information retrieval}

\keywords{Relevance Estimation,
Hybrid Search, Adaptive Retrieval}

\begin{abstract}

Advanced relevance models, such as those that use large language models (LLMs), provide highly accurate relevance estimations. However, their computational costs make them infeasible for processing large document corpora. To address this, retrieval systems often employ a telescoping approach, where computationally efficient but less precise lexical and semantic retrievers filter potential candidates for further ranking. However, this approach heavily depends on the quality of early-stage retrieval, which can potentially exclude relevant documents early in the process. In this work, we propose a novel paradigm for re-ranking called \textit{online relevance estimation} that continuously updates relevance estimates for a query throughout the ranking process. Instead of re-ranking a fixed set of top-k documents in a single step, online relevance estimation iteratively re-scores smaller subsets of the most promising documents while adjusting relevance scores for the remaining pool based on the estimations from the final model using an online bandit-based algorithm. This dynamic process mitigates the recall limitations of telescoping systems by re-prioritizing documents initially deemed less relevant by earlier stages---including those completely excluded by earlier-stage retrievers. We validate our approach on TREC benchmarks under two scenarios: hybrid retrieval and adaptive retrieval. Experimental results demonstrate that our method is sample-efficient and significantly improves recall, highlighting the effectiveness of our online relevance estimation framework for modern search systems.

\end{abstract}

\maketitle

\begin{figure}
    \centering
    \includegraphics[width=\linewidth]{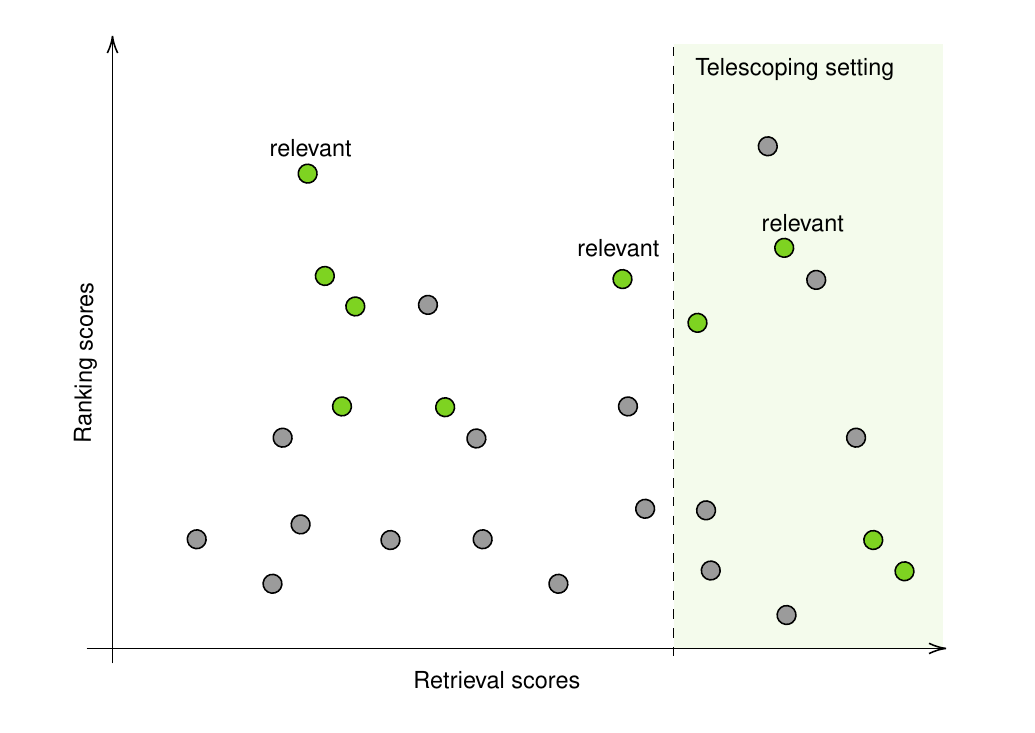}
    \caption{The distribution of retrieval and ranking scores of the retrieved documents. The green region represents the documents selected in the telescoping for ranking. The green documents are selected on the basis of online relevance estimation. The ground truth documents are explicitly labelled as ``relevant''.}
    \label{fig:intro-1}
\end{figure}

\section{Introduction}

Modern search engines are designed around the principle that only a small fraction of documents in a corpus are truly relevant to a given query, many of which can be identified using simple heuristics, such as lexical matching. Telescoping (or cascading) pipelines leverage this property to reduce the number of documents that need to be provided to more accurate (but more computationally expensive) relevance models such as those that use large language models (LLMs)~\cite{ma2024fine, pradeep2023rankvicuna, pradeep2023rankzephyr, Sun2023InstructionDM, sun2023chatgpt}. While this approach usually ensures that highly relevant documents appear at high ranks in the final result, the performance is ultimately limited by the recall of the early-stage retrievers.

The telescoping approach typically employs cost-effective retrievers such as those that rely on lexical~\cite{bm25,splade} or semantic~\cite{lin2021batch, karpukhin-etal-2020-dense} signals and efficient algorithms (such as BlocMaxWAND~\cite{ding2011faster} or HNSW~\cite{malkov2018efficient}) to perform initial candidate selection. To help ensure high recall, these systems are often combined into hybrid lexical-semantic ensembles~\cite{bruch2023analysis}, or extended using the nearest neighbors of the top documents with adaptive methods~\cite{macavaney2022adaptive}. These approaches achieve recall by ensuring broad coverage of potentially relevant documents. Subsequently, machine-learned rankers refine the top-k retrieved documents, optimizing precision-based measures with finer-grained relevance estimates.

Two major shortcomings limit telescoping pipelines. First, recall is inherently constrained by the quality of the initial retrieval stage, leading to the bounded-recall problem. Documents missed during this stage are irretrievably excluded from subsequent ranking, regardless of their relevance to the query. This over-reliance on early-stage retrievers undermines the system’s ability to recover highly relevant documents. For example, Figure~\ref{fig:intro-1} shows that relevant documents can be present beyond the top-k fold imposed in typical telescoping settings.
Second, documents from the early-stage retriever are processed in the order of their initial ranking scores, thereby filtering out documents that do not meet the re-ranking depth. Although the initial ranking may be a good initial prioritization of documents, we argue that processing the initial ranker's results order becomes suboptimal once the re-ranking model provides higher-quality relevance estimations. Although recent works~\cite{macavaney2022adaptive, kulkarni2023lexically, Kulkarni2024lexboost, rathee2024quam} have proposed adaptive retrieval to overcome the first problem, they still suffer from the second by relying on heuristics for prioritizing the candidate documents.

\begin{figure}
    \centering
    \includegraphics[width=\linewidth]{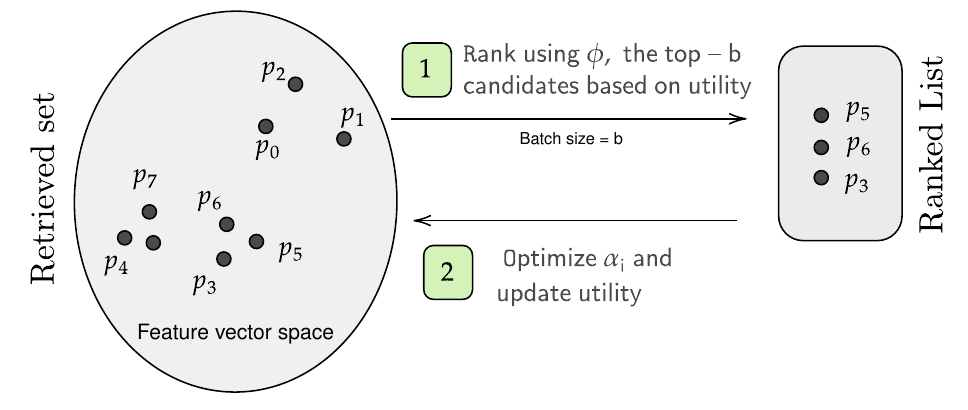}
    \caption{Schematic figure of the Online relevance estimation algorithm.}
    \label{fig:ORE}
\end{figure}

This work proposes a novel departure from the classical telescoping framework to address these limitations. Our approach, which we call \textit{online relevance estimation} (ORE), introduces a dynamic re-ranking paradigm that iteratively updates relevance estimates for the entire candidate pool throughout the ranking process. Instead of re-ranking a fixed top-k set of documents in one step, our method employs an iterative process that ranks smaller, high-potential subsets. The relevance scores of remaining documents are continuously refined based on the ranking outcomes, enabling previously overlooked documents to be revisited and reconsidered. This approach leverages an online bandit algorithm to optimize relevance estimation dynamically. Figure~\ref{fig:ORE} shows an overview of this process.

We validate our framework on TREC Deep Learning benchmarks under two practical retrieval scenarios: hybrid retrieval and adaptive retrieval. In hybrid retrieval, lexical and dense retrieval methods are fused to generate initial candidates, which are then re-ranked using cross-encoders. We demonstrate that online relevance estimation significantly improves recall by iteratively refining the rankings of a larger pool of documents. In the adaptive retrieval setting, which involves iterative ranking based on neighborhood exploration within a corpus graph, we show that our method surpasses existing approaches by explicitly estimating and updating candidate relevance scores. Unlike current adaptive retrieval methods, which focus on retrieving additional candidates, our approach integrates relevance estimation into the iterative process.

Experimental results highlight that our method is sample-efficient, offering \textbf{2$\times$} speedups over state-of-the-art, with the ORE component taking \textbf{10}$\times$ less time than expensive ranker calls. It also achieves substantial recall improvements, with upto \textbf{30.55 \%} gains on DL21 for adaptive retrieval and upto \textbf{14.12\%} gains on DL19 for hybrid retrieval with respect to the corresponding state-of-the-art. With respect to the standard telescoping baseline (BM25>>ranker), we achieve improvements of up to \textbf{58.53\%} on DL22. By bridging the retrieval and ranking stages, our online relevance estimation framework offers a scalable and effective solution to enhance the performance of search systems. %We release the code at~\url{https://anonymous.4open.science/r/ore}. 

\section{Related Work}
\label{sec:rel_work}

Recent advancements in document ranking have increasingly relied on complex rankers based on transformer models and, more recently, instruction-tuned models. These approaches have shown to be highly effective in delivering precise relevance estimates, particularly for nuanced ranking tasks. However, the computational cost of employing cross-encoders as rankers is substantial, with LLM-based rankers being even more expensive.
We contextualize our work into three parts, hybrid retrieval, adaptive retrieval, and other related ideas on online adaptation for rankings.

\subsection{Hybrid Retrieval and Telescoping}

Retrieval functions such as BM25 are generally designed to provide fast but less precise relevance estimates. In contrast, complex rankers, including cross-encoders and LLMs, offer far more accurate relevance assessments at the expense of significant computational resources. 
Due to this tradeoff, complex rankers are typically applied as final-stage ranking functions in a telescoping framework (also referred to as \textit{cascading} or \textit{multi-stage} ranking) \cite{matveeva2006high}. 
In this framework, an initial ranking is conducted using computationally inexpensive methods like BM25, and only a subset of top-ranked documents is passed to the final stage, where more expensive machine-learned models calculate the final ranking scores.
Consequently telescoping paradigm is widely adopted across a variety of domains where strict latency requirements are paramount: web search, e-commerce and live fact-checking systems.
Note that there is no restriction on what can be used as a retriever in the first stage. 
Historically lexical retrieval or BM25~\cite{bm25,splade, wang2011cascade} was mostly used as a retrieval function. 
In more modern search systems dense retrieval~\cite{lin2021batch, karpukhin-etal-2020-dense}, learned sparse~\cite{DBLP:conf/sigir/MacAvaneyN0TGF20b,splade}, and hybrid sparse-dense ensembles are used for first-stage retrieval~\cite{ cormack2009reciprocal,bruch2023analysis, chen2022out,wang2021bert}. 

However, telescoping suffers from a key limitation when the retrieval scores from the first stage do not accurately reflect the relevance of the documents. Retrieval scores are typically used to first-rank documents, and the top-k documents are selected for re-ranking based on their retrieval scores. Since this selection process is typically conducted in a single step, any failure to capture relevant documents in the top-k results can lead to poor recall, ultimately degrading precision in the final rankings. Furthermore, documents that are not passed to the re-ranking stage remain ranked solely according to their initial retrieval scores, which may not reflect their true relevance.

To improve the quality by improving recall, either higher retrieval depths are considered~\cite{leonhardt2022efficient,bruch2023analysis}, hybrid retrieval~\cite{bruch2023analysis,cormack2009reciprocal,hybrid-rank-lu2022zero} or query expansions techniques are employed~\cite{carpineto2012survey}.
Even if these approaches focus on a larger and a more varied retrieval set, the choice of documents to rank is still dependent on the retriever score.
Unlike these approaches, we dynamically update the relevance estimates for all retrieved candidates by iteratively ranking smaller batches of documents, resulting in improved recall. 

\subsection{Adaptive Retrieval}

The closest approaches to ours are the recently proposed Adaptive Retrieval (AR) methods introduced by~\citet{macavaney2022adaptive}. These methods operate on a corpus graph (constructed during an offline phase), which encodes document-document similarities based on lexical or semantic features.
Adaptive retrieval methods alternate between the initially retrieved results and the corpus graph neighborhoods of re-ranked documents to select a batch for re-ranking. These methods are fundamentally based on the Clustering Hypothesis~\cite{jardine1971use}, which assumes that relevant documents tend to cluster together in the feature space. In GAR~\cite{macavaney2022adaptive}, only the neighbors of previously ranked documents are explored. More recently, \textsc{Quam}~\cite{rathee2024quam} improved upon GAR by selecting documents based on their degree of relatedness to the re-ranked documents.

In contrast to GAR and \textsc{Quam}, which rely on cross-encoders for ranking, \citet{kulkarni2023lexically} proposed a method that uses bi-encoders to re-rank documents. Their approach selects only seed documents from the initial retrieved results and continues exploring the corpus graph neighborhood until the re-ranking budget is exhausted.
While adaptive retrieval methods dynamically schedule documents to the ranker, their alternating strategy is heuristic-driven and sample-inefficient. In contrast, our online relevance estimation framework generalizes and simplifies the adaptive retrieval paradigm, offering significant improvements in sample efficiency.

Partially related to our approach are ideas from online learning to rank~\cite{zoghi2017online,grotov2016online,li2019onlineltrs}, which learn the parameters of ranking models from user interaction data. However, our approach differs fundamentally from this line of work. Unlike these methods, we do not rely on direct user feedback or address challenges like prioritizing or de-biasing rank-sensitive clicks. Moreover, our framework operates on a significantly smaller feature space, allowing it to scale efficiently to large retrieval sizes compared to learning-to-rank models. Other similar works include~\citet{reddy2023inference} and \citet{DBLP:journals/corr/abs-2306-09657}, which learn a new query representation online during re-ranking. Unlike these methods, we use a bandit-based framework and continually refine query representations, thereby selecting better candidate documents at each inference step.

\section{Online Relevance Estimation}
\label{sec:method}

In document ranking, the task is to re-rank the top-$k$ documents retrieved from an initial retrieval stage using a more expensive ranker to produce the final ranked list. Typically, telescoping techniques (also referred to as cascading or multi-stage ranking) prioritize computational efficiency by pruning the document space with fast, less precise retrieval methods and then applying computationally expensive ranking functions (e.g., cross-encoders) to the remaining documents. However, such approaches suffer from low recall, as they rely solely on initial retrieval scores $\theta$ to schedule documents for re-ranking. Consequently, relevant documents with low retrieval scores may be overlooked, leading to reduced recall and precision in the final ranked list.

\subsection{Problem Definition}

The \cerberus{} framework is designed to estimate relevance scores for a large pool of retrieved or candidate documents $\mathcal{D}_q$ for a query $q$ such that the relative error between the estimated relevance scores (\textsc{EstRel}) and the cross-encoder scores (\(\phi\)) is minimized. Specifically, for a query \(q\) and a candidate document \(d \in \mathcal{D}_q\), the objective is to 
\begin{equation}
minimize \,\,\,\, \left| \phi(q, d) - {EstRel}(\vec{\alpha}, \vec{x}_d) \right|^2
\end{equation}
where \(\phi(q, d)\) represents the accurate relevance score provided by an expensive cross-encoder or ranker, and \textsc{EstRel}\((\vec{\alpha}, \vec{x}_d)\) is the estimated relevance score derived using simple document features \(\vec{x}_d\) and learnable parameters \(\vec{\alpha}\). The framework operates under the constraint of a strict budget \(m\), which limits the number of calls to the expensive ranker (\(\phi\)). This efficiency constraint ensures that only a subset of documents is scored directly using \(\phi\), while the relevance estimates for the remaining documents are derived from \textsc{EstRel}, which serves as a computationally inexpensive proxy for \(\phi\). The \cerberus{} framework presupposes that the cross-encoder \(\phi\) provides reliable relevance scores, which serve as ``ground truth'' for the estimation process.\footnote{Where \(\phi\) itself is an estimation of the true relevance of the document to the query.} By approximating \(\phi\) with simple, well-known relevance factors as characteristics (refer to Table~\ref{tab:features}), \cerberus{} aims to achieve an overall improvement in recall by effectively prioritizing highly relevant documents. This allows the framework to balance accuracy and efficiency, ensuring that the relevance estimates closely approximate \(\phi\) while adhering to the computational constraints imposed by the budget \(m\). As a result, \cerberus{} provides a scalable solution for large-scale document ranking tasks, achieving high-quality rankings while maintaining computational efficiency.

\algdef{SE}[DOWHILE]{Do}{doWhile}{\algorithmicdo}[1]{\algorithmicwhile\ #1}%

\begin{figure}[t]
\begin{algorithm}[H]
\caption{Online Relevance Estimation } 
\begin{algorithmic}
\Require Query $q$, initial retrieved pool $R_0$, batch size $b$, budget $c$,  number of batches to score $m$, features vector $\vec{x}_d$ for document $d$
\Ensure Scored pool $R_1$
\State $R_1 \gets \emptyset$ \Comment{Scored results}
\State $\mathcal{D}_q \gets R_0$ \Comment{candidate documents (Arms)}
\State $S \gets \emptyset$ \Comment{top ranked documents}
\State $\Vec{\alpha}_1 \leftarrow N(0,1)$ , $t \gets 1$
% \State $B \gets$ \Call{Score}{top $b$ from $A$, subject to $c$} \Comment{Using monoT5}
% \State $R_1 \gets R_1 \cup B$ \Comment{Add batch to results}
% \State $A \gets A \setminus B$ \Comment{Discard the batch from Arms}
% \State\HiLi $\Call{SetAff}{.} \gets 0$ \Comment{Initialize expected affinity scores}
\Do
%and $X_i = [x_1...x_n]$, $\forall d \in A $
% \State $X \gets [\vec{x_1}...\vec{x_l}]$ where $l=|A|$ \& $\vec{x_i}$- feature vector for document $d$  \Comment{Construct feature matrix features for documents}
%\frac{1}{m}  \sum_{i=1}^n \alpha_i  X_{i}]

  \State $\mathcal{D}_q \gets  \textsc{EstRel}(\Vec{\alpha}_t,\vec{x}_d) $ \quad $\forall d \in \mathcal{D}_q$  \Comment{Assign \textsc{EstRel} scores}
  \State $B \gets$ \Call{Select}{top $b$ from $\mathcal{D}_q$, subject to $c$} \Comment{using \textsc{EstRel}}
 
  %\State $S \gets$ \Call{Select}{top $s$ from $R_1$} \Comment{Select top $s$ ranked docs}
 % \State $A \gets  \Call{EstRel}{q,d,S} \quad \forall d \in A$  \Comment{Assign estimated relevance scores}

  % \State $U \gets top_u(\textsc{D2DAff}(A))$
  % \State $C \gets top_c(\textsc{Q2DAff}(A))$ \Comment{Create shortlists of candidates}

% \State\HiLii $\Call{SetAff}{d} \gets \Call{SetAff}{d} + \Call{SetAff}{d,R_1}, \quad \forall d \in F$ \Comment{\small{Update Exp. Aff. Scores using (\ref{eq:eaffm})}}
% \normalsize
% \State $F \gets \Call{InterSetAff}{q, d,S} \quad \forall d \in F$ \Comment{Assign interpolated set affinity scores}

%\State $B \gets$ \Call{Select}{top-$b$ from \{$U \cup C$\}, subject to $c$} 
% \State $B_2 \gets$ \Call{Select}{top-$l_2$, $EstRel$ scores from $A$, ; $l_1+l_2=b$}
\If{$|R_1| < m \cdot b$} 
    \State $B \gets$ \Call{Score}{$B$, subject to $c$} \Comment{e.g., monoT5}
    \State $\vec{\alpha}_{t+1} \gets \min_{\vec{\alpha}} E(\vec{\alpha}_t,q,d, \vec{x}_d)$ $\forall d \in B$
\Else 
\State  $B \gets$ \Call{LookUp}{\textsc{EstRel} scores} $ \forall d \in B$
\EndIf

% %\State $B \gets B_1 \cup  B_2 $  
\State $R_1 \gets R_1 \cup B$ \Comment{Add batch to results}
\State $\mathcal{D}_q \gets \mathcal{D}_q\setminus B$ \Comment{Discard batch from Arms}
\State $t \gets t+1$

\doWhile{$|R_1| < c$}
%\State $R_1 \gets R_1 \cup \Call{Backfill}{R_0, R_1}$ \Comment{Backfill remaining items}
\end{algorithmic}
\label{algo:cerberus}
\end{algorithm} 
\end{figure}

\begin{table*}
    \centering
            \caption{Description of different features used for calculating relevance estimates. These features can be divided into two levels of affinity taxonomy, \textsc{Q2DAff} and \textsc{D2DAff}.}
    \vspace{-1em}
    \begin{tabular}{cccccll}
        \toprule
        Feature & Notation & Taxonomy &\multicolumn{2}{c}{Source} & Description & \\
         &   & & Offline & Online  & \\
        \midrule
        $x_1$ &  $BM25(q,d)$ & \textsc{Q2DAff}& &   \cmark&Lexical similarity between query and document. \\
        $x_2$ & $TCT(q,d)$ & \textsc{Q2DAff} & &   \cmark &Semantic similarity between query and document. \\
        $x_3$ & $RM3(q^\prime, d)$&\textsc{D2DAff} & &\cmark &Lexical similarity between expanded query using RM3 and document. \\
        $x_4$ & $BM25(d,d^\prime)$ &\textsc{D2DAff} & \cmark & & Lexical similarity between pair of documents. \\
        $x_5$ & $TCT(d,d^\prime)$ &\textsc{D2DAff} & \cmark& & Semantic similarity between pair of documents. \\
        $x_6$ & $\textsc{Laff}(d,d^\prime)$ & \textsc{D2DAff}& \cmark & & Learnt affinity or similarity between pair of documents~\cite{rathee2024quam}. \\
        \bottomrule
    \end{tabular}
    \label{tab:features}
    %\vspace{-1.5em}
\end{table*}

\subsection{The \cerberus{}{} Framework}
\label{sec:framework}

The problem of relevance estimation in the \cerberus{} framework can be formulated as a top-$l$ arms selection problem in stochastic linear bandits~\cite{LUCB,chaudhuri2019pac}. In this formulation, the \textit{arms} correspond to candidate documents in the initial large retrieval or candidate document pool $\mathcal{D}_q$, the \textit{features vector} (\(\vec{x}_d\)) encode the properties of each document (as detailed in Table~\ref{tab:features}), and the \textit{rewards} represent the actual relevance scores (\(\phi(q, d)\)) obtained from the expensive ranker.
For a given query \(q\) and a candidate document \(d\), the estimated relevance score computed by \cerberus{} is expressed as:

\begin{equation}
\textsc{EstRel}(\vec{\alpha}, \vec{x}_d) = \vec{\alpha} \cdot \vec{x}_d^T,
    \label{eq:EstRel}
\end{equation}
where \(\vec{\alpha}\) represents the learnable parameters of the relevance estimation function. During training, the estimation error, which measures the discrepancy between the estimated relevance score (\textsc{EstRel}) and the actual relevance score (\(\phi\)), is minimized. The error is defined as:
\begin{equation}  
E(\vec{\alpha}; q, d, \vec{x}_d) = \frac{1}{2} \left| \phi(q, d) - \textsc{EstRel}\left(\vec{\alpha}, \vec{x}_d\right) \right|^2
\label{eq:error}
\end{equation}

While classical Multi-Arm Bandit (MAB) approaches iteratively update reward estimates by pulling arms until convergence, they typically require at least linear time in the number of arms per iteration. This makes them computationally impractical for large-scale document retrieval settings, where the candidate document pool can be vast. 
% Instead, the \cerberus{} framework is designed to address this scalability challenge by efficiently prioritizing and estimating relevance scores within a constrained computational budget.
Therefore, to ensure scalability, \cerberus{} constrains ranker calls (\(\phi\)) within a fixed budget \(m\). 
The framework performs parameter updates for a limited number of batches during re-ranking, learning the parameters \(\vec{\alpha}\) for the relevance estimator. For the remaining batches, the learned parameters \(\vec{\alpha}\) are used to estimate relevance scores for candidate documents. These estimated relevance scores are then used to add the candidate documents to the final ranked list, prioritizing based on their estimated relevance.
% This design allows \cerberus{} to approximate relevance scores effectively while significantly reducing the number of calls to the expensive ranker. By leveraging estimated relevance as a proxy for actual scores (\(\phi\)), \cerberus{} strikes a balance between computational efficiency and ranking accuracy. As a result, the framework enables a more sample-efficient approach to online relevance estimation, making it suitable for large-scale document retrieval tasks.

\subsection{Query Processing using \cerberus{}}
\label{sec:algo}

Algorithm~\ref{algo:cerberus} provides an overview of the \cerberus{} procedure. Let \(q\) denote the query, \(R_0\) represent the initial pool of retrieved documents, and \(R_1\) the final re-ranked pool of documents, which is initially empty. Let \(S\) be the set of top \(s\) documents from $R_1$ that have been re-ranked so far (initially empty), \(b\) the batch size, and \(c\) the re-ranking budget. The candidate document pool is denoted as \(\mathcal{D}_q\), which is initialized with the results retrieved during the first stage (depending on whether the retrieval setup is Hybrid or Adaptive). For each document \(d \in \mathcal{D}_q\), let \(\vec{x}_d\) denote its feature vector. 
Each document \(d \in \mathcal{D}_q\) is assigned an estimated relevance score, \textsc{EstRel}, computed using Equation~\ref{eq:EstRel} with an initial parameter vector \(\vec{\alpha}_1\), which is sampled from a normal distribution (\(\vec{\alpha}_1 \sim N(0,1)\)). The \textsc{EstRel} score quantifies the utility or perceived importance of a document in \(\mathcal{D}_q\).

The \cerberus{} procedure begins by selecting a batch \(B\) of the top \(b\) documents from \(\mathcal{D}_q\), based on their \textsc{EstRel} scores. These documents are scored using the expensive ranker \(\phi\) (e.g., MonoT5~\cite{nogueira2020document}), and the re-ranked documents are added to \(R_1\). Following this, \cerberus{} updates \(\mathcal{D}_q\) by either exploring the neighborhood graph (in Adaptive Retrieval) or expanding the retrieval depth (in Hybrid Retrieval) to include additional candidate documents.

To prioritize documents for ranking, the framework recomputes \textsc{EstRel} scores for all documents in \(\mathcal{D}_q\) using Equation~\ref{eq:EstRel}. A new batch \(B\) of the top \(b\) documents, based on their updated \textsc{EstRel} scores, is selected for ranking. The selected batch is scored using the expensive ranker \(\phi\), and the parameters \(\vec{\alpha}\) of the relevance estimator are updated by minimizing the estimation error as defined in Equation~\ref{eq:error}. These updated parameters are then used to recompute \textsc{EstRel} scores for the remaining documents in \(\mathcal{D}_q\).

The expensive ranker \(\phi\) is used until the condition \(|R_1| < m \cdot b\) is satisfied, where \(m\) represents the maximum number of batches that can be scored using \(\phi\). For subsequent documents, the learned parameters \(\vec{\alpha}\) are reused to estimate relevance scores, and batches are selected based on their \textsc{EstRel} scores. These selected documents are then added to \(R_1\). The process of updating \textsc{EstRel} scores and selecting batches continues iteratively until the condition \(|R_1| < c\) is met, where \(c\) is the re-ranking budget.
The intuition behind scoring only a subset of documents lies in approximating the relevance of a candidate document \(d \in \mathcal{D}_q\) using a learned combination of its features \(\vec{x}_d\). By prioritizing and scoring a limited number of batches with \(\phi\), the learned parameters \(\vec{\alpha}\) enable accurate relevance estimation for the remaining documents. This approach eliminates the need for scoring all documents with the ranker, providing significant efficiency gains while maintaining competitive performance, as demonstrated in Section~\ref{sec:estimated rel}.
\section{Estimated Relevance in \cerberus{}}
\label{sec:estimated rel}

 Relying only on retrieval scores (\textsc{Q2DAff}) would lead to omission of documents which might be relevant, as shown in Figure \ref{fig:intro-1}. However, these documents may have closer proximity to documents already deemed relevant as measured by document-document similarity/affinity (\textsc{D2DAff}). If we compute the affinity of the document with respect to a set of documents, it is termed as \textsc{D2SETAff}. 
 
 Hence, the choice of features used in \cerberus{} is the cornerstone of quality in online relevance estimation. A summary of features employed in \cerberus{} in different setups (hybrid and adaptive) is as shown in Table \ref{tab:features}. Apart from \textsc{Q2DAff} scores, we also capture the proximity of the document to a small set of documents already deemed relevant by the expensive ranker. The intuition follows from the explore-exploit paradigm of linear stochastic bandits. In the current setting, our goal is to allow for the balance between prioritizing documents with high retrieval scores (exploitation) or provisioning selection of documents which have closer proximity to highly relevant documents despite it's lower retrieval scores (exploration). Note that, \cerberus{} is not limited to only the features in the Table \ref{tab:features}. The design of \cerberus{} algorithm makes it flexible towards the addition of new features. 

For a given query $q$, let $R_0$ be the initial retrieved results with lexical ($x_1$) or semantic ($x_2$) query-document similarities, \textsc{Q2DAff}. and $R_1$ be the results after re-ranking. Let $G_c$ be the corpus graph. The corpus graph $G_c$ encodes lexical ($x_4=BM25(d,d^\prime)$) or semantic ($x_5=TCT(d,d^\prime)$) document-document similarities, \textsc{D2DAff}. Let $G_a$ be the learnt affinity graph, proposed in \quam{}~\cite{rathee2024quam}, which encodes learnt affinity, \laff{} scores ($x_6$).

\subsection{Hybrid Retrieval using~\cerberus{}}
Hybrid retrieval usually entails employing multiple lexical (BM25) and dense (TCT) retrievers for a high retrieval depth, followed by rank fusion to merge the retrieved lists. These approaches then usually cap the merged results to a lower retrieval depth, ignoring other potentially relevant documents with lower retrieval scores. However, \cerberus{} promotes exploration by constructing a candidate pool of documents from the entire merged list. In the hybrid retrieval setup, our goal is to prioritize not only documents with high retrieval scores (\textsc{Q2Daff}) but also balance the exploration of documents that are in close proximity to documents (\textsc{D2DAff}) already deemed highly relevant. Hence, we carefully select the features from Table \ref{tab:features} reflective of this philosophy

\[\Call{Q2DAff}{q,d} = \alpha_1 * BM25(q,d) + \alpha_2 * TCT(q,d)\] $\forall d \in \mathcal{D}_q$, where $\alpha_1, \alpha_2 \in \vec{\alpha}$.

For \textsc{D2DAff} features in the hybrid retrieval context, we employ both lexical (RM3 i.e., $x_3$) and semantic scores ($TCT(d,d^\prime)$, i.e., $x_5$). It is critical to note that these \textsc{D2DAff} scores are employed to compute \textsc{D2SetAff} scores, which measure the proximity of a candidate document to a set of highly relevant documents. These highly relevant documents are selected as top-s documents that have already been scored so far from $R_1$.
\[ \textsc{D2SetAff}=\alpha_3 * RM3(q^\prime,d) + \alpha_4 * \frac{\sum_{d^\prime \in S} (\phi(q,d^\prime) *TCT(d,d^\prime))}{|S|}\]
where $\alpha_3, \alpha_4 \in \vec{\alpha}$ and $q^\prime$ is the expanded query by using $RM3$ expansion over top re-ranked documents so far in $R_1$. Note, we simply look up the score of $\phi(q,d^\prime)$ since $d^\prime$ is already re-ranked using ranker $\phi$. Mapping this to Equation~\ref{eq:EstRel}, \textsc{EstRel} is computed using $\vec{\alpha}=[\alpha_1,\alpha_2,\alpha_3,\alpha_4]$ and $\vec{x}_d$=[$x_1$, $x_2$, $x_3$, $x_5$]. The parameters in $\vec{\alpha}$ are learnt using the mechanism described in Section 3.1 and Algorithm \ref{algo:cerberus}. 

%\mpara{Query-Document Affinity.} 
%Given a candidate document $d$, it's affinity to the original query $q$ is defined as the degree of relevance indicated by the scores from first-stage retrieval.  For instance, if BM25 is employed for first-stage retrieval, query-document affinity,  \textsc{Q2DAff}

\subsection{Adaptive Retrieval using \cerberus{}}

We adopt a similar philosophy for document prioritization in the adaptive retrieval setup. However, adaptive retrieval is a bit more involved, as the candidate pool $\mathcal{D}_q$ is not static and expands with the addition of neighbors of top-scored documents. Hence, the relevance estimation for the candidate documents is linear in terms of number of documents (arms). Hence, we draw inspiration from top-$l$ arm selection in linear stochastic bandits like LUCB \cite{LUCB} and GIFA \cite{reda2021top} which maintain multiple sets such as : 1) arms with high reward estimates and 2) arms with low reward estimates to balance exploration. However, these approaches still sample actual rewards for one arm from each of these lists rendering them computationally infeasible for a large candidate pool.
Hence, we maintain two shortlists which represent 1) documents (arms) with high \textsc{Q2Daff} scores, denoted by $U \subset \mathcal{D}_q$, and 2) documents with high \textsc{D2SetAff} scores, denoted by $V\subset \mathcal{D}_q$. The intuition is that since \textsc{EstRel} primarily depends on balancing between documents with high \textsc{Q2Daff} and documents with high  \textsc{D2SetAff} scores maintaining shortlists based on these measures help reduce the expanding candidate space and also reduce the impact of documents with noisy estimates.

In the adaptive setting, $\Call{Q2DAff}{q,d} = BM25(q,d)$. Given a document affinity graph $G_a$, the document-document affinity is given by: $\Call{D2DAff}{d,d^\prime } = G_a(d,d^\prime) $ where  $G_a(d,d^\prime)$ is the edge weight or edge affinity between the source document $d$ and its neighbor $d^\prime$ in the corresponding graph. Note that, we lookup \textsc{Q2Daff}(q,d) and \textsc{D2DAff}(q,d) and compute \textsc{EstRel} $\forall$ $d \in U \cup V$ thereby providing an efficient relevance estimation mechanism.
Our goal is to primarily balance the exploitation paradigm with the exploration. The exploitation primarily entails selecting documents that have high affinity to the query. Whereas, the exploration paradigm entails scheduling neighbors that may not have high affinity to the query but are closely related to multiple documents deemed to be highly relevant to the query. To accomplish this, we compute the affinity of the candidate document to the ranked set of documents $S$, denoted as \textsc{SetD2DAff} and defined as
\begin{equation}
\Call{D2SetAff}{d, S} =\frac{\sum_{d^\prime \in S \cap N_d} (\textsc{D2DAff}(d, d^\prime))} {|S \cap N_d|}
\label{eq:SetD2Daff}
\end{equation}
where $N_d = \Call{Neighbours}{d, G_a}$ is the set of neighbors of document $d$ in the learnt affinity graph $G_a$. The estimated relevance ($\textsc{EstRel}$) of the candidate document $d$ to the given query $q$ can be better estimated using an average of the relevance (score given by the ranker $\phi$) of documents from $S$ in its neighborhood that are already deemed to be highly relevant to the query. Hence we also include this new feature in adaptive retrieval as it naturally fits into the neighborhood-based retrieval philosophy of this setup.

% where $P(Rel(d^{\prime}))$ encodes the estimated relevance distribution induced by a relevance scorer (e.g., MonoT5~\cite{nogueira2020document}) model.
% $P(R(d^{\prime}))$ can be estimated in multiple ways. 
% We let $P(Rel(d^{\prime})) =\frac{e^{\phi\left(q,d\right)}}{\sum_{d^{\prime} \in S} e^{\phi\left(q,d^{\prime} \right)}}$.

% \mpara{Utility of the document.} To balance the query-document affinity and document-document set affinity, we interpolate \textsc{Q2DAff} and \textsc{SetD2DAff}.
% The interpolated measure is the estimated utility of a candidate document $d$ defined as
% \begin{equation}
% \Call{U}{q,d,S}= \alpha \cdot \Call{Q2Daff}{q,d}  + \beta \cdot \Call{SetD2DAff}{d, S}
% \label{eq:utility}
% \end{equation}
% where $\beta$, s.t.  $0\leq\beta\leq 1$, is a parameter for interpolation. 

\begin{equation}
x_7 = \frac{\sum_{d^\prime \in S \cap N_d}{\Call{Score}{q,d^\prime}}}{|S \cap N_d |}
\end{equation}

Hence, the features can be combined in the following form for adaptive retrieval: 
\vspace{-0.1cm}
\[\alpha_1 * \textsc{Q2DAff}(q,d) + \alpha_2 * \textsc{D2SetAff}(q,d)  + \alpha_3 * x_7\]
\vspace{-0.5cm}
\begin{equation}
  \Call{Score}{q,d^\prime} =  \begin{cases} 
       \phi{(q, d^\prime )} + \psi(q, d^\prime ) \ ; \ if \ d^\prime \ is \ scored \ using \ \phi \\
        \Call{EstRel}{\vec{\alpha},\vec{x}_d} \ ; \ otherwise \\
   \end{cases}
\label{eq:score}
\end{equation}
where $\psi$ is a dual encoder\footnote{We use inexpensive dual encoder, TAS-B~\cite{hofstatter2021efficiently} for better numerical stability.} and $S^\prime$ is set of top $s$ documents in previous iteration. We look up the scores from $\phi$, since the documents are already in $R_1$ ($d^\prime\in S^\prime \subseteq R_1$). Mapping this to Equation~\ref{eq:EstRel}, \textsc{EstRel} is computed using $\vec{\alpha}=[\alpha_1,\alpha_2,\alpha_3]$ and $\vec{x}_d$=[$x_1$, $x_6$, $x_7$]. %These  parameters in $\vec{\alpha}$ are learned using the mechanism described in Section 3.1 and Algorithm \ref{algo:cerberus}.

% The above measure is termed \textit{estimated relevance}, $\textsc{EstRel}$, and is used as a proxy for the importance of a document. The documents with high $\textsc{EstRel}$ scores are scheduled for ranking. This score balances the notion of affinity to query and affinity to highly relevant documents. It allows for principled exploration of useful documents that may not have a high affinity to query but have high document affinity, $\textsc{D2DAff}$ to the neighbors of re-ranked documents in the set $S$, while balancing exploitation of documents with high affinity to query, $\textsc{Q2DAff}$.

Note that all the above computations for hybrid or adaptive retrieval setups are vectorized and computed for a batch of documents at a time. We present at the document level for ease of understanding. Also, note that $ \mathcal{D}_q$ get updated after scoring each batch $B$ with the neighbors in $G_a$ of each document $d \in B$, i.e., $ \mathcal{D}_q$ $\gets$ $\mathcal{D}_q$  $ \cup$ \Call{Neighbors}{$d$,$G_a$}, but we maintain shortlists as discussed earlier.  

% \subsection{Offline Policy for Multi-Arm-Bandit framework}

\section{Experimental Setup}
\label{sec:experiment}

In this work, we demonstrate the effectiveness of online relevance estimation in two commonly used recall-improving scenarios: \textit{hybrid retrieval} and \textit{adaptive retrieval}. To evaluate our approach, we address the following research questions:

% In this work, we address the following research questions:
\begin{enumerate}
\item[\bf RQ1:] How effective is  \cerberus{} compared to existing approaches for hybrid and adaptive retrieval setups?
\item[\bf RQ2:] How helpful is the utility (estimated relevance) in prioritizing documents for retrieval?
\item[\bf RQ3:] How  efficient is \cerberus{} compared to existing approaches for adaptive retrieval? 
\item[\bf RQ4:] How much time does estimated relevance take compared to expensive ranker calls?  

\end{enumerate}

\subsection{Datasets and Measures}
We perform experiments on the MSMARCO passage corpus~\cite{bajaj2016ms} (with 8.8 M passages) and validate our approach on the TREC Deep Learning 2019 (DL19) and 2020 (DL20)~\cite{craswell2021trec} test sets. The DL19 set has 43 queries and DL20 has 54 queries. Further, we use the MSMARCO passage-v2 corpus~\cite{bajaj2018msmarcohumangenerated} (with 138.4 M passages) and evaluate on TREC DL21 and DL22 test sets. The DL21 has 53 queries and DL22 has 76 queries. We use the de-duplicated MSMARCO-passage-v2 corpus and both DL21 and DL22 qrels. We measure the ranking performance by nDCG@c, and retrieval by recall@c at different re-ranking budgets $c\in\{50,100,1000\}$. We re-use the BM25-based and TCT-based corpus graphs created in \gar{}. 
%Specifically, we re-use the BM25-based (encodes lexical similarities) and TCT-based (encodes semantic similarities) corpus graphs.  

\subsection{Retrieval and Ranking Models}
We mainly use lexical and semantic first-stage retrievers. For lexical retrieval, we use BM25~\cite{bm25}. We use a Terrier~\cite{ounis2005terrier} index of the MSMARCO passage corpus. While for semantic retrieval, we use TCT~\cite{lin2021batch} which is based on the TCT-Colbert model, and use the TCT-ColBERT-HNP\footnote{\url{https://huggingface.co/castorini/tct\_colbert-v2-hnp-msmarco}} model for encoding queries and documents. We retrieve documents based on the budget (in the adaptive retrieval setting) or retrieval depth (in the hybrid retrieval setting). We also use RM3~\cite{abdul2004umass} query expansion leveraging BM25 index.

We use the MonoT5-base model~\cite{nogueira2020document} (in short MonoT5) as the ranker model which is fine-tuned on the MSMARCO corpus. MonoT5 is based on cross encoder setting which takes the query and document together as input and predicts the relevance score. We also use MonoT5 as a retriever on the MSMARCO passage corpus by scoring all documents exhaustively of a query. Also, we do ablation using the fine-tuned pointwise LLM ranker called RankLLaMA~\cite{ma2024fine}, which is built upon LLaMA-2-7B\footnote{\url{https://huggingface.co/meta-llama/Llama-2-7b-hf}} and trained for ranking the top documents from the RepLLaMA retriever. 
%\todo{rankLLAMA is not discussed}

\subsection{Baselines and Implementation}
\label{sec:scenarios-setup}

To compare the effectiveness of our proposed method, we use re-ranking, hybrid, and adaptive retrieval baselines. We use a standard telescoping re-ranking baseline, retriever followed by ranker, by re-ranking top retrieved documents based on the re-ranking budget $c$. We denote this ranking baseline by BM25>>MonoT5. 

\noindent \textbf{Hybrid Retrieval.} For hybrid retrieval, we use two, BM25 and TCT, retrievers for the first stage and retriever 1000 documents exhaustively. We apply Reciprocal Rank Fusion~\cite{cormack2009reciprocal} (RRF) over these two rankings and take the top $c$ (budget) documents based on their reciprocal rank scores. We also use Convex Combination~\cite{bruch2023analysis,wang2021bert} (CC) of scores given by BM25 and TCT retriever with interpolation parameter $\alpha$ is set to 0.5~\footnote{As~\cite{bruch2023analysis} mentioned that the CC methods are sensitive to $\alpha$,  we follow the insight from ~\cite{wang2021bert} that $\alpha=0.5$ works best for lexical and semantic interpolation for the MSMARCO corpus.}.

\noindent \textbf{Adaptive Retrieval.} For adaptive retrieval, we use mainly \gar{}~\cite{macavaney2022adaptive} and \quam{}~\cite{rathee2024quam}. Both \gar{} and \quam{} alternate between first-stage results and neighborhood graph and prepare the batch of documents for reranking. For both \gar{} and \quam{}, we use BM25 and TCT-based corpus graphs with 16 neighbors.  The type of corpus is indicated in subscript, for example, \gar{} with BM25 based corpus graph is denoted by \sys{BM25}. We use the official implementation to reproduce these baselines.

% \gar{} mainly uses \textsc{Q2DAff} and \textsc{D2DAff} as features and select documents alternatively based on these features. On the other hand, \quam{} proposed and used the \textsc{SetD2DAff}, an additional feature to \gar{}, which measures the utility of a candidate document to the already re-ranked documents. 

\subsection{Hyperparameters and Tuning}
For our experiments, we use re-ranking budget $c \in \{50,100,1000\}$, and batch size is set to $16$. We mainly use the corpus graphs with $16$ neighbors. We use DL19 set as a validation set for tuning hyperparameters and DL20, DL21, and DL22 as test sets. For $RM3$, we set \textit{fb\_docs} to 5 and \textit{fb\_terms} to 10, and the \textit{original\_query\_weight} to 0.3. We set $|S|=10$ for all budgets in hybrid retrieval. We set $|U|=35, |V|=25$ for different re-ranking budgets $c$. For adaptive retrieval setup, we set the size of set $S$ to calculate the \textsc{D2SetAff} depending upon the budget. For budget $c$ of 50, 100, and 1000, we set $|S|$ to 10, 25, and 150 respectively. All of our experiments are done on NVIDIA H100 GPU with 96 GB of RAM. 
\section{Experimental Results}
\label{sec:results}

We extensively evaluate the effectiveness, and efficiency of \cerberus{} in hybrid and adaptive retrieval scenarios.
Note that for hybrid retrieval, we consider a fixed large retrieval depth constructed by the union of the documents retrieved using lexical matching and semantic similarity as discussed in~\ref{sec:scenarios-setup}. The initial results were prioritized by result fusion, with the \textit{top-k} results being scored by the re-ranker (MonoT5).

\begin{table*}
    \centering
    \caption{Effectiveness comparison of \cerberus{} with hybrid and adaptive retrieval methods on TREC DL19 and DL20 test sets. Significant improvements using paired t-test, $p<0.05$, with Bonferroni correction, over CC, RRF, baseline (BM25>>MonoT5),  \gar{}, and \quam{} are marked with $B$, $C$, $R$, $G$ and $Q$ respectively. The best scores are highlighted in bold. }
    \vspace{-1em}
    {\small
     \setlength{\tabcolsep}{2.5pt}
    \begin{tabular}{llrrrrrrrrr}
        \toprule
        \multicolumn{1}{l}{}&\multicolumn{1}{c}{}&\multicolumn{3}{c}{$c=50$}&\multicolumn{3}{c}{$c=100$}&\multicolumn{3}{c}{$c=1000$} \\
\cmidrule(lr){3-5}\cmidrule(lr){6-8}\cmidrule(lr){9-11}
Dataset&Pipeline  & nDCG@10 & nDCG@c & Recall@c & nDCG@10 & nDCG@c  & Recall@c & nDCG@10 & nDCG@c & Recall@c\\
    \midrule
     & \bf\textsc{Exhaustive Retrieval} & \\
     \cmidrule{2-11}
\multirow{11}{*}{\textbf{DL19}} \cellcolor{white}& MonoT5   & 0.672 & 0.625& 0.512  & 0.672 & 0.611  & 0.599 &0.672 & 0.691 & 0.834  \\
\cmidrule{2-11}
 & \bf\textsc{Hybrid Retrieval}: (BM25 \& TCT) & \\
   &  RRF>>MonoT5 [R] & \bf0.735& 0.658 & 0.513 & 0.729 & 0.664 & 0.637 &\bf0.703 &0.740 &0.879 \\
   &  CC>>MonoT5 [C] & 0.729& 0.650 & 0.489 & \bf0.730 & 0.650 & 0.626 &0.698 &0.738 &0.878 \\
   &  \cerberus{}  & 0.734 & ${}^{RC}$\bf{0.683}&${}^{RC}$\bf0.558 & 0.721 & ${}^{RC}$\bf0.688 & ${}^{RC}$\bf0.675 & \bf 0.703 &\bf0.741 &\bf0.882\\

   % &  \cerberus{} (BM25, TCT, $score_{diff}$) & \textbf{0.} & \textbf{0.}&\textbf{0.} & \textbf{} & \textbf{0.671} & \textbf{0.653}  & 0.702 & 0.739 & 0.878\\
   
    \cmidrule{2-11}
      &   \bf\textsc{Adaptive Retrieval} & \\

    \cmidrule{2-11}
 \cellcolor{white} & BM25>>MonoT5 [B] &0.681  & 0.541 &0.389 & 0.699 &0.563   &0.488&0.719 &0.697 &0.755  \\
     %  &  \bf\textsc{Alternating} & &  &  &  & \\
     &     \idnt w/ \sys{BM25} [G]  & 0.689  & 0.565 & 0.417 & 0.716 &0.594 &0.539 &0.727 &0.742 &0.836 \\
       % &  \idnt w/ \sys{BM25}+Laff  &0.705  &0.596  &0.444  & 0.734 &0.643 &0.593 &0.735 &0.761 &0.870 \\
        &  \idnt w/ \quamsys{BM25} [Q]  & 0.698 &0.597  & 0.460 &\bf0.729 &0.639& 0.594 &\bf0.742 &\bf0.770 &0.874 \\

       %   \cmidrule{2-11}
       % &  \idnt w/ \sys{TCT}   & 0.712  & 0.608 & 0.472 &  &  &\\
       %  & \idnt w/ \sys{TCT}+Laff   &0.706  &0.601  &0.456 &  & & \\
       %   & \idnt w/ \quamsys{TCT}  & 0.693 &0.599  & 0.466 &  &  &\\
      %   \cmidrule{2-11}
      % &  \bf\textsc{Non-Alternating} & &  &  &  & \\
      %  &  \idnt w/ \sys{BM25}   & 0.683 & 0.546 & 0.401 &0.689 &0.551 &0.496\\
      %  & \idnt w/ \sys{BM25}+Laff  &0.711  &0.615  &0.465 & 0.739&0.655 &0.626 \\
      %   & \idnt w/ \quamsys{BM25}  & 0.715 &0.628  & 0.491 &0.717 &0.632 &0.590\\
        %   &   \idnt w/ CERBERUS (half)  & 0. 706 &0.636  & 0.509 &0.725  & 0.653  & 0.640 & & & \\
        % &   \idnt w/ CERBERUS   & \textbf{0. 727}& \textbf{0.657}  & \textbf{0.532} &\textbf{0.733}  & \textbf{0.672}  & \textbf{0.649} & & & \\
          % &   \idnt w/ CERBERUS++ (half)  & 0.687 & \textbf{0.619}& \textbf{0.494} & 0.683 & 0.621 & \textbf{0.598}  \\

        &   \idnt w/ \cerberussys{BM25}   & 0.698 & ${}^{GQ}_{B}$\textbf{0.640}& ${}^{GQ}_{B}$\textbf{0.509} & 0.711 & ${}^{G}_{B}$\textbf{0.653} & ${}^{G}_ {B}$\textbf{0.619} & 0.723 & ${}^{}_{B}$0.759 & ${}^{}_{B}$\textbf{0.874} \\

 % \cline{2-11}
 %         &   \idnt w/ \sys{TCT} [G]  & 0.712 & 0.608 & 0.472 & 0.732 &0.642  &0.592  & 0.741 & 0.764 &0.859 \\
 %        &   \idnt w/ \quamsys{TCT} [Q]   &0.692 & 0.600 & 0.466 & 0.722 &0.635  &0.595  &0.742 &0.767 &0.880 \\
 %        &   \idnt w/ \cerberussys{TCT}  &0.695& \textbf{0.637} & \textbf{0.501}  & 0.720 &\textbf{0.654} & \textbf{0.612}  & 0.736 & 0.763 & \textbf{0.884} \\
      %  \cmidrule{2-11}
       % &  \idnt w/ \sys{TCT}   &0.716  & 0.620 & 0.486 &  &  &\\
       %  & \idnt w/ \sys{TCT}+Laff  &0.716  &0.617  &0.478 &  &  & \\
       %   & \idnt w/ \quamsys{TCT}  & 0.703 &0.621  & 0.484 &  &   &\\

        \midrule
             & \bf\textsc{Exhaustive Retrieval} & \\
            \cmidrule{2-11}    
    \cellcolor{white} \multirow{11}{*}{\textbf{DL20}}  & MonoT5 &0.649  & 0.592 &0.576 & 0.649 &0.593 &0.670 &0.649 &0.682 &0.852 \\
        \cmidrule{2-11}    
     & \bf\textsc{Hybrid Retrieval}: (BM25 \& TCT)& \\
   &  RRF>>MonoT5 [R]& \bf0.721& 0.655 & 0.633 & 0.707 & 0.659 & 0.725 &0.676 &0.727 &0.885 \\
   &  CC>>MonoT5 [C]& 0.718& 0.654 & 0.632 & \bf0.709 & 0.660 & 0.721 &\bf0.681 &0.727 &0.884 \\
   &  \cerberus{}  & 0.720 & ${}^{}$\bf0.674&${}^{}$\bf0.658 & 0.702 & ${}^{RC}$\bf0.683 & ${}^{RC}$\bf0.759 & 0.676 & ${}^{C}$\bf0.731 & ${}^{C}$\bf0.892 \\
      % &  w/  \cerberus{} (BM25+TCT+$score_{diff}$) & \textbf{0.714} & \textbf{0.658}&\textbf{0.647} & \textbf{0.701} & \textbf{0.662} & \textbf{0.733} & 0.677 & 0.727 & 0.883 \\
       \cmidrule{2-11}
      &   \bf\textsc{Adaptive Retrieval} & \\

    \cmidrule{2-11}
          \cellcolor{white}  & BM25>>MonoT5 [B] &0.676 & 0.559 &0.478 &0.685 &0.581 &0.584  &\bf0.720 &0.711  &0.807 \\
      % &  \bf\textsc{Alternating} & &  &  &  & \\
     &     \idnt w/ \sys{BM25} [G] & 0.690 & 0.577 & 0.496 & 0.703 &0.607 &0.617 &0.714 &0.750 &0.884 \\
       % &  \idnt w/ \sys{BM25}+Laff  &0.707 &0.607  &0.548  &0.716 &0.643 &0.665 &0.708 &0.753 &0.898 \\
        &  \idnt w/ \quamsys{BM25} [Q]  & \bf0.714 &0.615  & 0.553 &\bf0.717 &0.652  &0.678 &0.709 &0.756 &0.901 \\

       %   \cmidrule{2-11}
       % &  \idnt w/ \sys{TCT}   & 0. & 0. & 0. &  &  &\\
       %  & \idnt w/ \sys{TCT}+Laff   &0.  &0.  &0. &  & & \\
       %   & \idnt w/ \quamsys{TCT}  & 0. &0.  & 0. &  &  &\\
       %  \cmidrule{2-11}
      % & \bf \textsc{Non-Alternating} & &  &  &  & \\
     %   &  \idnt w/ \sys{BM25}   & 0.670 & 0.534 & 0.459 &0.702& 0.587& 0.601\\
     %   & \idnt w/ \sys{BM25}+Laff  &0.706  &0.597  &0.538 & 0.726& 0.648 &0.691 \\
     %    & \idnt w/ \quamsys{BM25}  & 0.713 &0.616 & 0.543 & 0.719 &0.648&0.684 \\
          % &   \idnt w/ CERBERUS (half)  & 0.696  &0.613  & 0.577 &0.699  & 0.654  & 0.705 & 0.  & & \\
          %           &   \idnt w/ CERBERUS   & 0. 694 &\textbf{0.642}  & \textbf{0.612} &0.707  & \textbf{0.663}  & \textbf{0.721} & & & \\

              % &   \idnt w/ CERBERUS++ (half)  & 0.672 &0.606  & \textbf{0.575} &0.668  & 0.621  & 0.684 & & & \\
              
        &   \idnt w/ \cerberussys{BM25}   & 0.684& ${}^{G}_{B}$\textbf{0.621}  & ${}^{G}_{B}$\bf{0.583} &0.681  & ${}^{G}_{B}$0.651  & ${}^{G}_{B}$\textbf{0.705} & 0.700& \bf0.757 & ${}^{}_{B}$0.892 \\
        % &   \idnt w/ \sys{TCT} [G]  &0.728 & 0.620 & 0.566 & 0.716 &0.648 & 0.683  &0.706 & 0.749 &  0.893\\
        % &   \idnt w/ \quamsys{TCT} [Q]   &0.717 & 0.618 & 0.572 & 0.713 &0.653 & 0.701  & 0.706 & 0.752 & 0.901 \\
        % &   \idnt w/ \cerberussys{TCT}   &0.680 & \textbf{0.637} & \textbf{0.598} & 0. &0.649 & \textbf{0.706}  &  \\
       % &  \idnt w/ \sys{TCT}   &0.  & 0. & 0. &  &  &\\
       %  & \idnt w/ \sys{TCT}+Laff  &0.  &0.  &0. &  &  & \\
       %   & \idnt w/ \quamsys{TCT}  & 0. &0.  & 0. &  &   &\\
      \bottomrule     
    \end{tabular}
    }
    \vspace{-1.em}
    \label{tab:main}
\end{table*}
\subsection{Effectiveness of \cerberus{}}

In the first experiment, we want to evaluate the effectiveness of online relevance estimation over the telescoping strategy used over standard, hybrid, and adaptive retrieval. To address \textbf{RQ1}, we evaluate \cerberus{} on TREC-DL 2019 and 2020 datasets, comparing its performance to state-of-the-art methods in hybrid and adaptive retrieval setups in Tables~\ref{tab:main} and~\ref{tab:msmarco-v2}.
Firstly, online relevance estimation outperforms baseline ranking performance in telescoping settings i.e., BM25>>MonoT5 (up to \textbf{58.53\%} on DL22 at budget $c=100$). 

\subsubsection{Hybrid Retrieval}

We now turn our attention to hybrid retrieval. As expected, we confirm that both RRF>>MonoT5 and CC>>\\MonoT5 convincingly outperformed BM25>>MonoT5 at all retrieval depths.
This is because using hybrid retrieval balances the complementary lexical and semantic signals.
We find that \cerberus{} further improves beyond this baseline, achieving statistically significant performance gains over both RRF and CC.
For instance, from Table \ref{tab:msmarco-v2}, we observe substantial gains, where \cerberus{} outperforms CC by \textbf{11.74 \%} and RRF by \textbf{17.12\%} for Recall@100 on DL21. We also observe that \cerberus{} improves Recall@100 on DL22 by \textbf{7.46 \%}, when compared to CC  and by \textbf{14.09\%} when compared to RRF. Further on DL19, Recall@50 improves from $0.489$ to $0.558$ (an improvement of \textbf{14.11\%}) and from $0.513$ to $0.558$ (an improvement of \textbf{8.9\%}) in CC and RRF respectively. 
Similar trends are observed across different retrieval budgets, with \cerberus{} delivering consistent gains.

These improvements can be primarily attributed to \cerberus{}'s online relevance estimation capability, which prioritizes documents dynamically based on the current estimate of the relevance.  
Unlike fusion-based methods that select the \textit{top-k} merged documents based on a one-shot fusion score and ignore others, \cerberus{} captures potentially relevant documents with low initial retrieval scores by re-prioritizing them for scoring based on new ranking evidence. 
Our Multi-Arm Bandits-based online estimation procedure trades off exploration (scheduling low-ranked documents) with exploitation (scoring top-ranked documents), thereby effectively learning the tradeoffs between relevance factors modelled as features. Given our small feature space and linear classifier, \cerberus{} can perform this reprioritization efficiently. Future work could explore the trade-offs of extended feature sets.

\subsubsection{Adaptive Retrieval}
We also compare \cerberus{} to state-of-the-art adaptive retrieval methods, including \gar{} and \quam{}. 
Unlike the hybrid setting, where the retrieval set is fixed, in the adaptive retrieval setting, we adaptively explore the retrieved document space.
Our results indicate that \cerberus{} outperforms these approaches across various retrieval budgets, with significant gains at lower budgets. For example, on DL21, we observe that \cerberus{} advances Recall@50 to \textbf{0.406} providing gains of up to \textbf{30.55\%} over \quam{} and up to \textbf{22.66\%} over \gar{}.
On TREC-DL 2019 Recall@50 increases from $0.460$ (\quam{}) and $0.417$ (\gar{}) to $0.509$ (\textbf{10.65\%} and \textbf{22.06\%}, respectively).
These gains arise from the principled document selection strategy employed by \cerberus{}. 
Existing methods like \gar{} and \quam{} alternate between first-stage retrieval results and neighborhood lists. 
We believe that this alternating strategy was proposed in the spirit of ensuring the robustness of results and might be sometimes less sample efficient.
For example, the algorithm is forced to schedule documents from the retrieved list to be ranked even though the retrieval scores are low and indicate low relevance.
 \cerberus{} departs from the alternating scheduling strategy by re-estimating document utility over all the candidate documents -- from the retrieved results or the neighborhood graph.
 % The estimation is typicalonline through an inexpensive relevance estimation technique (\textsc{EstRel}, detailed in Section \ref{sec:estimated rel}). 
 This approach enables the balanced exploitation of documents retrieved in the first stage and the exploration of related documents identified in their neighborhoods. 
As a result, \cerberus{} prioritizes relevant documents at each iteration that may have low initial retrieval scores but high estimated utility.
Surprisingly, \cerberus{} also outperforms the exhaustive retrieval pipeline, which uses an expensive scorer to evaluate all documents in the corpus without first-stage retrieval. This highlights the effectiveness of \cerberus{}'s utility estimation in reducing noise and focusing on potentially relevant documents.

\noindent \textbf{Insight 1:}
\cerberus{} achieves high recall in both hybrid and adaptive retrieval settings by dynamically learning to prioritize documents through an inexpensive estimation of relevance scores.

\subsection{Significance and Quality of Estimated Utility}
\label{sec:rq-2}

While the overall performance demonstrates that the proposed utility estimation aids in prioritizing documents, it does not provide insights into its absolute quality or its ability to serve as a reliable proxy for the expensive ranker’s scores. To answer \textbf{RQ2}, we evaluate the quality of estimated scores in the hybrid and adaptive retrieval setups.

% Since the utility computed for documents online (\textsc{EstRel}) forms the cornerstone of \cerberus{}, to answer \textbf{RQ2}, we analyze the quality of scores estimated online. 

\begin{table}
    \centering
    \caption{Effectiveness comparison* of \cerberus{} with hybrid and adaptive retrieval methods on TREC DL21 and DL22 test sets. The letter in subscript or superscript shows significant improvements (using paired t-test, $p<0.05$, with Bonferroni correction) over the corresponding baseline. The best score for each pipeline is highlighted in bold. }
    \vspace{-1em}
    {\small
     \setlength{\tabcolsep}{2.5pt}
    \begin{tabular}{llrrrr}
        \toprule
        \multicolumn{1}{l}{}&\multicolumn{1}{c}{}&\multicolumn{2}{c}{$c=50$}&\multicolumn{2}{c}{$c=100$} \\
\cmidrule(lr){3-4}\cmidrule(lr){5-6}
Dataset &Pipeline  &   nDCG@c & Recall@c &   nDCG@c  & Recall@c \\
    \midrule
  \multirow{12}{*}{\textbf{DL21}} & \bf\textsc{Hybrid } & \\
 \cmidrule{2-6}
   &  RRF>>MonoT5 [R] & 0.576 & 0.401  & 0.558 & 0.520  \\
   &  CC>>MonoT5 [C] & 0.584 & 0.419  & 0.569 & 0.545  \\
   &  \cerberus{}   & ${}^{R}$\bf{0.604}&${}^{R}$\bf0.444  & ${}^{RC}$\bf0.609 & ${}^{RC}$\bf0.609\\

    \cmidrule{2-6}
      &   \bf\textsc{Adaptive} & \\

    \cmidrule{2-6}
 \cellcolor{white} & BM25>>MonoT5 [B]   & 0.436 &0.242  &0.433 &0.331  \\
     &     \idnt w/ \sys{BM25} [G]    & 0.457 & 0.290  &0.465 &0.414  \\
        &  \idnt w/ \quamsys{BM25} [Q]   &0.478  & 0.310  &\bf0.499& 0.454  \\

        &   \idnt w/ \cerberussys{BM25}  & ${}^{GQ}_{B}$\textbf{0.503}& ${}^{GQ}_{B}$\textbf{0.364}  & ${}^{}_{B}$0.481 & ${}^{G}_{B}$ \bf0.463 \\
           
    \cmidrule{3-6}  
     &     \idnt w/ \sys{TCT} [G]   & 0.502 & 0.331  &\bf0.520&0.489  \\
        &  \idnt w/ \quamsys{TCT} [Q]   &0.491  & 0.311  &0.518& 0.477  \\

        &   \idnt w/ \cerberussys{TCT}    & ${}^{GQ}_{B}$\textbf{0.532}& ${}^{GQ}_{B}$\textbf{0.406} & ${}^{}_{B}$0.512 & ${}^{}_{B}$ \bf0.502 \\
                    \midrule

     \multirow{12}{*}{\textbf{DL22}} & \bf\textsc{Hybrid } & \\
      \cmidrule{2-6}
   &  RRF>>MonoT5 [R]& 0.452 & 0.260  & 0.430 & 0.341 \\
   &  CC>>MonoT5 [C]& 0.459 & 0.278 & 0.433 & 0.362\\
   &  \cerberus{}   & ${}^{RC}$\bf0.481&${}^{R}$\bf0.297  & ${}^{RC}$\bf0.459 & ${}^{RC}$\bf0.389  \\
      % &  w/  \cerberus{} (BM25+TCT+$score_{diff}$) & \textbf{0.714} & \textbf{0.658}&\textbf{0.647} & \textbf{0.701} & \textbf{0.662} & \textbf{0.733} & 0.677 & 0.727 & 0.883 \\
       \cmidrule{2-6}
      &   \bf\textsc{Adaptive} & \\

    \cmidrule{2-6}
          & BM25>>MonoT5 [B]  & 0.290 &0.115  &0.275 &0.164   \\
      % &  \bf\textsc{Alternating} & &  &  &  & \\
     &     \idnt w/ \sys{BM25} [G]  & 0.287 & 0.121 &0.290 &0.191  \\
       % &  \idnt w/ \sys{BM25}+Laff  &0.707 &0.607  &0.548  &0.716 &0.643 &0.665 &0.708 &0.753 &0.898 \\
        &  \idnt w/ \quamsys{BM25} [Q]   &\bf0.308  & 0.135  &\bf0.303  &\bf0.196  \\
              
        &   \idnt w/ \cerberussys{BM25}   & ${}^{}_{}$ 0.292  & ${}^{}_{}$\bf{0.137}   & ${}^{}_{}$0.284  & 0.195 \\
         \cline{3-6}
     &     \idnt w/ \sys{TCT} [G]    & 0.329 & 0.157  &\bf0.348&0.256  \\
        &  \idnt w/ \quamsys{TCT} [Q]   &0.329  & 0.155  &0.334& 0.237  \\

        &   \idnt w/ \cerberussys{TCT}    & ${}^{GQ}_{B}$\textbf{0.364}& ${}^{GQ}_{B}$\textbf{0.206}  & ${}^{}_{B}$0.342 & ${}^{}_{B}$ \bf0.260 \\
         \midrule
           
    \end{tabular}
    }
    \begin{minipage}{\columnwidth}
    \footnotesize{*We omit nDCG@10 measure because of space constraints and our focus is more on retrieval. Additionally, prior works find that nDCG@10 value saturates quickly during re-ranking, while evaluation at lower depths are able to further distinguish systems~\cite{plaid_repro}.}
\end{minipage}
    \vspace{-2em}
    \label{tab:msmarco-v2}
\end{table}

\subsubsection{Hybrid Retrieval}

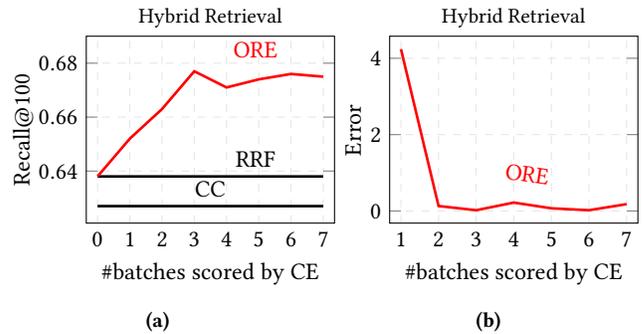
\begin{figure}
    \centering
    \begin{subfigure}{0.48\columnwidth}
        \centering
 \begin{tikzpicture}
		\begin{axis}[
			width =  1.2\linewidth,
			height =1.\linewidth,
			major x tick style = transparent,
			grid = major,
		    grid style = {dashed, gray!20},
			xlabel = {\#batches scored by CE},
			ylabel = {Recall@100},
			title={\small{Hybrid Retrieval}},
            title style={yshift=-1.5ex}, % Shift title closer to the plot
            symbolic x coords={0,1,2,3,4,5,6,7},
            xtick={0, 1,2,3,4,5,6,7},
            xtick distance=20,
            enlarge x limits=0.05,
            xlabel near ticks,
            ylabel near ticks,
            % every axis x label/.style={at={(0.5, -0.08)},anchor=near ticklabel},
            ymax=0.69,
            every axis y label/.style={at={(-0.18, 0.5)},rotate=90,anchor=near ticklabel},
			]

	% \addplot [color=black, style= dashed, line width = 1.0pt] table [x index=0, y index=1, col sep = comma] {plots/ablation/crossenc_vs_perf_hybrid.txt}
 %        node[pos=0.5, anchor=west, yshift=7pt] {BM25};
			
	\addplot [color=black, line width = 1.0pt] table [x index=0, y index=2, col sep = comma] {plots/ablation/crossenc_vs_perf_hybrid.txt}node[pos=0.7, sloped, above] {RRF};

   	\addplot [color=black, line width = 1.0pt] table [x index=0, y index=3, col sep = comma] {plots/ablation/crossenc_vs_perf_hybrid.txt}
    node[pos=0.5, sloped, above] {CC};

   	\addplot [color=red, line width = 1.0pt] table [x index=0, y index=4, col sep = comma] {plots/ablation/crossenc_vs_perf_hybrid.txt}
    node[pos=0.7, yshift=5pt, above] {\cerberus{}};
   
    \end{axis}
    \end{tikzpicture}    
    \caption{} 
    \label{fig:hybrid_cross_enc_vs_perf}
    \end{subfigure} 
    \hspace{0.02\columnwidth}
   % \hfill
    \begin{subfigure}{0.48\columnwidth}
        \centering
    \begin{tikzpicture}
		\begin{axis}[
			width =  1.2\linewidth,
			height =1.0\linewidth,
			major x tick style = transparent,
			grid = major,
		    grid style = {dashed, gray!20},
			xlabel = {\#batches scored by CE},
			ylabel = {Error},
			title={\small{Hybrid Retrieval}},
            title style={yshift=-1.5ex}, % Shift title closer to the plot
            symbolic x coords={1,2,3,4,5,6,7},
            xtick={1,2,3,4,5,6,7},
            xtick distance=20,
            enlarge x limits=0.05,
            enlarge y limits=0.08,
            xlabel near ticks,
            ylabel near ticks,
            %ymax=0.8,
            % every axis x label/.style={at={(0.5, -0.08)},anchor=near ticklabel},
            every axis y label/.style={at={(-0.08, 0.5)},rotate=90,anchor=near ticklabel},
			]

	% \addplot [color=black, style= dashed, line width = 1.0pt] table [x index=0, y index=1, col sep = comma] {plots/ablation/intro.txt}
 %        node[pos=0.5, anchor=west, yshift=7pt] {BM25};
			
	% \addplot [color=black, line width = 1.0pt] table [x index=0, y index=2, col sep = comma] {plots/ablation/intro.txt}node[pos=0.7, sloped, above] {\sys{}};

 %   	\addplot [color=black, line width = 1.0pt] table [x index=0, y index=3, col sep = comma] {plots/ablation/intro.txt}
 %    node[pos=0.5, sloped, above] {\quamsys{}};

   	\addplot [color=red, line width = 1.0pt] table [x index=0, y index=4, col sep = comma] {plots/ablation/cross_enc_vs_error.txt}
    node[pos=0.7, yshift=5pt, sloped, above] {\cerberus{}};
   
        \end{axis}
    \end{tikzpicture}    
    \caption{ } 
    \label{fig:hybrid_error}
    \end{subfigure} 
    \vspace{-2em}
    \caption{Recall (left) and estimation error (right) comparison for hybrid retrieval setting on the \textbf{TREC DL19} dataset when the number of batches of scored by cross-encoder (CE) varies for \cerberus{} for ranking budget of 100 and batch of size 16.}
    \label{fig:hybrid}    
    \Description{}
    \vspace{-1.0em}
\end{figure}

 We analyze the error between the estimated relevance (\textsc{EstRel}) and the actual relevance scores from the ranker for various ranker call budgets ($m$). Here, the budget for ranker calls $m$ represents the number of batches of documents scored by the ranker, with $m \cdot b \leq c$. The error for $c=100$ and $b=16$ across $m={1, \dots, 7}$ is shown in Figure \ref{fig:hybrid_error}.

\newcommand{\gt}[1]{\tiny{(\tikz\draw[green!60!black, fill=green!60!black] (0,0) -- (0.10,0) -- (0.05,0.10) -- cycle;{#1\%}})}

\newcommand{\rt}[1]{\tiny{(
    \tikz\draw[red, fill=red] (0,0) -- (0.10,0) -- (0.05,-0.10) -- cycle;%
    {#1\%})}}

\begin{table*}
\centering
\caption{Mean re-ranking latency per query (in ms) at different re-ranking budgets using MonoT5 and RankLLaMA rerankers when the first-stage retrieval of different budgets ($c$) is done using BM25. The number of batches re-ranked by ranker \cerberus{} is enclosed in braces. }
\vspace{-1.5em}
\label{tab:latency}
\setlength{\tabcolsep}{4.5pt}
\begin{tabular}{clllccc|lllccc}
\toprule
& \multicolumn{6}{c|}{MonoT5}& \multicolumn{6}{c}{RankLLaMA}  \\
\cmidrule{2-7}\cmidrule{8-13}
&  \multicolumn{3}{c}{time (ms/query)}& \multicolumn{3}{c|}{Recall@c} &\multicolumn{3}{c}{time (ms/query)}& \multicolumn{3}{c}{Recall@c}  \\
\cmidrule(lr){2-4} \cmidrule(lr){5-7} \cmidrule(lr){8-10} \cmidrule(lr){11-13} 

c   & \sys{} & \quamsys{}  & \cerberussys{} &\sys{}  & \quamsys{} & \cerberussys{}  & \sys{} & \quamsys{}  & \cerberussys{} &\sys{}  & \quamsys{} & \cerberussys{}\\
\midrule
50    & 179.92  & 173.21  & 125.91(2)    & 0.417 & 0.460  & \textbf{0.500} &6269.77   &6027.12   &  3925.54(2)   & 0.421 & 0.449 & \textbf{0.492} \\
100    & 356.53  &328.82  & 272.04(4)    &  0.539 &0.594   & 0.594 & 12746.55  &12074.67  & 7830.41(4)    & 0.542  & \textbf{0.600}  & \textbf{0.600} \\
250     &877.19 &816.90 &599.24(8)   & 0.692 & 0.745  &0.715 & 32312.97 & 30539.78 &  16092.16(8) & 0.684 &0.761  &  0.719\\
% 500    &   &   &   &   & 0. & 0.  &  \\
% 750  &   &    &   &   & 0. & 0.  &  \\
\multirow{2}{*}{1000}    &  3418.98 & 3219.45  & 1848.26(8)  & 0.836 &0.874  & 0.827  & 127617.45  & 120885.39   & 16327.25(8) & 0.854 &0.881 & 0.829 \\
&   &   &  2188.29(16)  &  &  & 0.841  &   &   & 31939.78(16) & &  & 0.853 \\

\bottomrule
\end{tabular}
\vspace{-1.em}
\end{table*}

At $m=1$, the error is high because the parameters used for estimating utility are initialized randomly, resulting in poor relevance approximations. However, as $m$ increases, the utility estimates improve significantly. 
For instance, at $m=2$, only 32 samples are scored, but the learned parameters enable a sharp reduction in error, closely approximating the actual relevance scores. As more samples are scored with increasing $m$, the error continues to decline steadily, reflecting better utility estimation.

This trend is further supported by Figure \ref{fig:hybrid_cross_enc_vs_perf}, which shows that when only 16 documents are scored, \cerberus{} already outperforms traditional hybrid retrieval methods, such as RRF and CC. By estimating high-quality utility scores for the remaining documents, \cerberus{} achieves superior performance even with minimal ranker calls.
\subsubsection{Adaptive Retrieval}
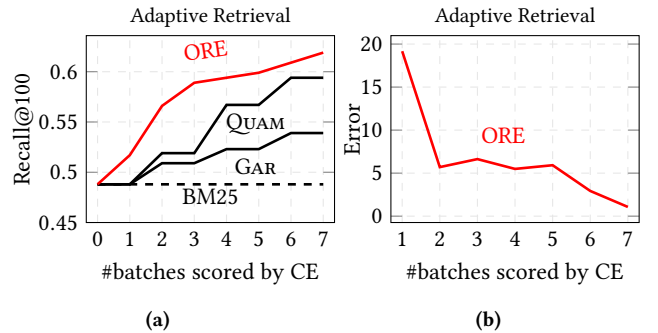
\begin{figure}
    \centering
    \begin{subfigure}{0.48\columnwidth}
        \centering
    \begin{tikzpicture}
		\begin{axis}[
			width =  1.2\linewidth,
			height =1.0\linewidth,
			major x tick style = transparent,
			grid = major,
		    grid style = {dashed, gray!20},
			xlabel = {\#batches scored by CE},
			ylabel = {Recall@100},
			title={\small{Adaptive Retrieval}},
            title style={yshift=-1.5ex}, % Shift title closer to the plot
            symbolic x coords={0, 1,2,3,4,5,6,7},
            xtick={0,1,2,3,4,5,6,7},
            xtick distance=20,
            enlarge x limits=0.05,
            xlabel near ticks,
            ylabel near ticks,
            ymin=0.45,
            % every axis x label/.style={at={(0.5, -0.08)},anchor=near ticklabel},
            every axis y label/.style={at={(-0.18, 0.5)},rotate=90,anchor=near ticklabel},
			]

	\addplot [color=black, style= dashed, line width = 1.0pt] table [x index=0, y index=1, col sep = comma] {plots/ablation/intro.txt}
        node[pos=0.5, anchor=west, yshift=2pt, below] {BM25};
			
	\addplot [color=black, line width = 1.0pt] table [x index=0, y index=2, col sep = comma] {plots/ablation/intro.txt}node[pos=0.7, sloped, below] {\sys{}};

   	\addplot [color=black, line width = 1.0pt] table [x index=0, y index=3, col sep = comma] {plots/ablation/intro.txt}
    node[pos=0.7, sloped, below] {\quamsys{}};

   	\addplot [color=red, line width = 1.0pt] table [x index=0, y index=4, col sep = comma] {plots/ablation/intro.txt}
    node[pos=0.5, sloped, yshift=5pt, above] {\cerberus{}};
   
    \end{axis}
    \end{tikzpicture}    
    \caption{ } 
    \label{fig:adaptive_cross_enc_call_vs_perf}
    \end{subfigure} 
    \hspace{0.02\columnwidth}
    %\hfill
    \begin{subfigure}{0.48\columnwidth}
        \centering
    \begin{tikzpicture}
		\begin{axis}[
			width =  1.2\linewidth,
			height = 1.0\linewidth,
			major x tick style = transparent,
			grid = major,
		    grid style = {dashed, gray!20},
			xlabel = {\#batches scored by CE},
			ylabel = {Error},
			title={\small{Adaptive Retrieval}},
            title style={yshift=-1.5ex}, % Shift title closer to the plot
            symbolic x coords={1,2,3,4,5,6,7},
            xtick={1,2,3,4,5,6,7},
            xtick distance=20,
            enlarge x limits=0.05,
            xlabel near ticks,
            ylabel near ticks,
            % every axis x label/.style={at={(0.5, -0.08)},anchor=near ticklabel},
            every axis y label/.style={at={(-0.09, 0.5)},rotate=90,anchor=near ticklabel},
			]

   	\addplot [color=red, line width = 1.0pt] table [x index=0, y index=4, col sep = comma] {plots/ablation/ar_cross_enc_vs_error.txt}
    node[pos=0.7, yshift=5pt, above] {\cerberus{}};
   
        \end{axis}
    \end{tikzpicture}  

    \caption{ } 
    \label{fig:adaptive_error}
    \end{subfigure} 
    \vspace{-2em}
    \caption{Recall (left) and estimation error (right) comparison on the \textbf{TREC DL19} dataset for adaptive retrieval, for ranking budget of 100 and batch of size 16.}
    \label{fig:ar_plots}    
    \vspace{-1.8em}
    \Description{}
\end{figure}

For adaptive retrieval, we analyze the error in estimated utility, as illustrated in Figure \ref{fig:adaptive_error}. The results reveal a trend similar to the hybrid retrieval setup: the error decreases gradually as the cross-encoder budget ($m$) increases, with a sharp decline observed at the maximum budget ($m=7$), where more samples are scored. Additionally, we examine the relationship between the ranker budget and Recall@100 for TREC-DL 2019 (DL19) in Figure \ref{fig:adaptive_cross_enc_call_vs_perf}. Across all ranker budgets ($m$), \cerberus{} consistently outperforms state-of-the-art adaptive retrieval methods, such as \gar{} and \quam{}. Notably, even at $m=1$, \cerberus{} demonstrates superior performance by using its utility estimates as a proxy for ranking documents, avoiding frequent calls to the expensive ranker. This result highlights the high quality of the estimated utility scores, which enable \cerberus{} to prioritize relevant documents through principled exploration.

 % when the number of batches scored by cross-encoder varies for \cerberus{} 

The observed improvement is attributed to the heuristic-based document selection strategies employed by \gar{} and \quam{}, which alternate between the initial ranked list and the neighborhood of scored documents. At lower ranker budgets, these methods score only a limited number of documents and backfill the remaining slots with scores from the initial retrieval results. In contrast, \cerberus{} employs a learned utility estimator that performs principled exploration. It dynamically prioritizes documents from the initial retrieval results and their neighborhoods in each batch until the budget $c$ is fully utilized. 

\noindent \textbf{Insight 2:} Online relevance/utility estimation of \cerberus{} works well across hybrid and adaptive retrieval settings and closely approximates actual relevance estimates from the ranker.

%\vspace{-1.5em}
\subsection{Computational Efficiency of \cerberus{}}

\begin{figure}
    \centering
    \begin{subfigure}{0.42\columnwidth}
        \centering
    \begin{tikzpicture}
    \begin{axis}[
    width =  1.4\linewidth,
	height =1.2\linewidth,
	major x tick style = transparent,
	grid = major,
	grid style = {dashed, gray!20},
    ybar=0.9,
    bar width=6.0, % Set the width of the bars here
    enlargelimits=0.15,
    title style={yshift=-1.5ex}, % Shift title closer to the plot
    legend style={at={(0.20,.88)},
       anchor=north,legend columns=-1, font=\tiny},
    ylabel={},
    xlabel={Recall@100},
    title={\small{time (ms/query)}},
    every axis y label/.style={at={(-0.00, 0.5)},rotate=90,anchor=near ticklabel},
    symbolic x coords={0.51,0.56, 0.59},
    xtick={0.51,0.56, 0.59},
    nodes near coords={},
    nodes near coords align={vertical},
    ylabel near ticks,
    xlabel near ticks,
    yticklabel style = {font=\small,xshift=0.5ex},
    xticklabel style = {font=\small,yshift=0.5ex},
    legend columns=1,
    legend entries={\sys{},\quamsys{},\cerberus{} },
    enlarge x limits=0.25]

\addplot[fill=gray!50] coordinates {(0.51, 68 ) (0.56,238) (0.59,276)};
\addplot[fill=blue!20] coordinates {(0.51, 65) (0.56, 130 ) (0.59,194 )};
\addplot[fill=red!30] coordinates {(0.51, 63) (0.56,102 ) (0.59, 178)};

\end{axis}
\end{tikzpicture}
    \caption{} 
        \label{fig:latency_adaptive_bars}
    \end{subfigure} 
    %\hfill
    \hspace{0.5cm}
    \begin{subfigure}{0.42\columnwidth}
        \centering
    \begin{tikzpicture}
    \begin{axis}[
    width =  1.4\linewidth,
	height =1.2\linewidth,
	major x tick style = transparent,
	grid = major,
	grid style = {dashed, gray!20},
    ybar=0.9,
    bar width=6.0, % Set the width of the bars here
    enlargelimits=0.15,
    legend style={at={(0.20,0.88)},
       anchor=north,legend columns=1, font=\tiny},
    ylabel={},
    xlabel={Budget $c$},
    title={\small{time (ms/query)}},
    title style={yshift=-1.5ex}, % Shift title closer to the plot
    every axis y label/.style={at={(0.15, 0.5)},rotate=90,anchor=near ticklabel},
    symbolic x coords={50,100, 250},
    xtick={50,100, 250},
    nodes near coords={},
    nodes near coords align={vertical},
    ylabel near ticks,
    xlabel near ticks,
    yticklabel style = {font=\small,xshift=0.5ex},
    xticklabel style = {font=\small,yshift=0.5ex},
    legend columns=1,
    legend entries={CE ,Fit, Lookup},
    enlarge x limits=0.25]

\addplot[fill=black!60] coordinates {(50, 144.7 ) (100,259) (250,601.7)};
\addplot[fill=purple!20] coordinates {(50, 4.5) (100, 9.18 ) (250,22.9 )};
\addplot[fill=yellow!30] coordinates {(50, 4.9) (100,11.25 ) (250, 32.2)};

\end{axis}
\end{tikzpicture}
    \caption{} 
        \label{fig:latency_cereberus_comp}
    \end{subfigure} 
    \label{fig:latency}  
    \vspace{-1.em}
    \caption{Computational efficiency of our proposed methods \cerberus{} in comparison to adaptive retrieval approaches (left) and overheads from different components (right) during online relevance estimation.}
    \Description{}
\end{figure}
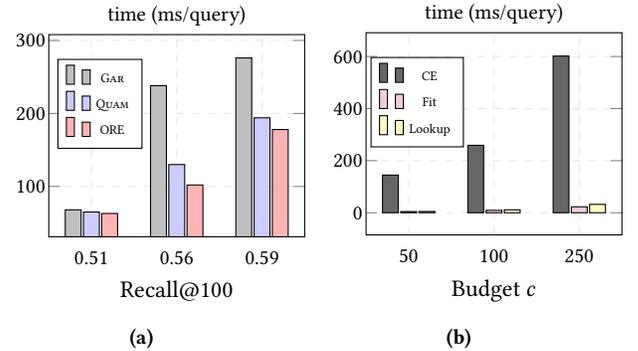

 To address \textbf{RQ3}, we demonstrate the latency and sample efficiency gains provided by \cerberus{} over \gar{} and \quam{} in the adaptive retrieval setting. In Figure~\ref{fig:latency_adaptive_bars}, we present the time taken by \cerberus{} and contemporary adaptive retrieval methods to achieve similar recall performance. Specifically, to reach a Recall@100 of $0.56$, \cerberus{} requires only 2 cross-encoder calls (\textit{102 ms/query}), providing a speedup of \textbf{2$\times$} compared to \gar{}, which takes 8 calls (\textit{238 ms/query}). This highlights \cerberus{}’s ability to achieve higher recall with fewer scored samples due to its efficient online relevance estimation.

From Table~\ref{tab:latency}, we observe that \cerberus{} consistently outperforms existing adaptive retrieval methods in terms of both latency and Recall@c when using an expensive ranker such as RankLLaMa. On average, \cerberus{} delivers speedups of \textbf{2$\times$–3$\times$} and, in certain scenarios, achieves up to \textbf{9$\times$} speedups for $c=1000$ across different budgets compared to \gar{} and \quam{}. These improvements primarily stem from the sample-efficient nature of \cerberus{}, which requires fewer scored samples to estimate utility scores for the remaining documents. These utility scores serve as reliable proxies for actual relevance scores, significantly reducing need for costly ranker calls.

Further, to answer \textbf{RQ4} we provide a breakdown of the time taken by individual components of \cerberus{} in Figure \ref{fig:latency_cereberus_comp} for $c=\{50,100,250\}$, batch size $b=16$ on DL19. These components are namely, the expensive ranker calls (denoted by CE), feature construction (denoted by Lookup), and parameter updates (i.e., fitting of $\vec{\alpha}$ parameters, denoted by Fit). 
As we discussed earlier, during re-ranking, the expensive ranker contributes the most in the latency overhead. We observe similar insights here. For example, at budget $c=250$, the total time for re-ranking is around 657 ms/query, out of which the cross encoder (CE) contributes around 92\% (602 ms/query) of the time.  
The feature lookup takes only 32.2 ms/query (\textbf{$18 \times$} less time) compared to 601.7 ms/query for ranker calls. Similarly, the time taken to learn and update $\alpha$ parameters takes only 22.9 ms/query. Hence, the core component for relevance estimation (parameter fitting and feature lookup) takes \textbf{10 $\times$} less time than ranker calls at per query level. 

\noindent{\bf{Insight 3:}} \cerberus{} is sample efficient when compared to state-of-the-art adaptive retrieval methods. It requires fewer documents scored by the expensive ranker on average. It provides speedups of upto \textbf{2$\times$} for standard rankers like MonoT5 and upto \textbf{9 $\times$} for more expensive LLM-based rankers like RankLLaMa.

\vspace{-0.6em}
\section{Conclusion}

In this work we introduce a novel paradigm of dynamically ranking  retrieved documents by using online relevance estimation.
We propose a departure from the progressive filtering approach popularized by the telescoping method that only ranks documents with high retrieval scores ignoring other retrieved documents.
Instead, we propose to dynamically keep relevance estimates for every retrieved document based on a small set of features based on well-known relevance factors.
These estimates are refined dynamically by incorporating ranking scores encountered during the ranking process.
Our experiments suggest that our framework of online relevance estimation is flexible, general, and easy to use in many retrieval settings.
Our experiments over four TREC-DL datasets in the hybrid and adaptive retrieval settings clearly show that basic instantiations of online relevance estimation are quite effective and outperform other telescoping and adaptive retrieval baselines.

\bibliographystyle{ACM-Reference-Format}
\balance
\bibliography{bib}

% \appendix

\end{document}